\documentstyle[12pt]{article}

\newcommand{\be}{\begin{equation}}
\newcommand{\ee}{\end{equation}}

\newcommand{\p}{\partial}
\newcommand{\Lbd}{\Lambda}
\newcommand{\lbd}{\lambda}

\newcommand{\g}{\gamma}

\date{\today}
\title{
 Instanton solutions in the problem
 of wrinkled flame fronts dynamics.
 }
 \author{Dmitri Volchenkov $ {}^{1,2} $  and Ricardo Lima ${}^{1}$ }

\begin{document}
\thispagestyle{empty}
\noindent
 \maketitle

{\small   \em $ {}^{1} $ Centre de Physique Theorique, 13288 Marseille
  Cedex 09, Luminy Case 907, France }

{\small  \em $ {}^{2} $Institute of Physics, Saint-Petersburg
State University, Peterhof, Ulyanovskaya 1,
198904, Russia}

{\small   \em volchen@cptsg2.univ-mrs.fr, lima@cpt.univ-mrs.fr}

PACS number(s): 47.10.+g, 47.27.-i,82.40.Py, 47.54.+r, 05.40.+j,

\begin{abstract}

The statistics of wrinkling  flame front is investigated
by the quantum field theory methods.
We dwell on the WKB approximation in the
functional integral which is  analogous to the Wyld functional
integral in turbulence. The main contribution to statistics
is due to a coupled field-force configuration. This configuration
is related to a kink between metastable exact pole solutions
of the Syvashinsky equation. These kinks are responsible
for both the formation of new cusps and the rapid
power-law acceleration of the mean flame front. The
problem of asymptotic stability of the solutions is discussed.

\end{abstract}

\section{Introduction}

It had been shown in \cite{1} that  under a weakly nonlinear approximation,
the dynamics of a wrinkled planar flame front is governed by a nonlinear
partial differential equation (PDE)
\be
\frac{\p \Phi}{\p t}=\frac 12 U_{b}\left(\frac{\p\Phi}{\p x}\right)^{2}+
D_{M}\frac{\p^2 \Phi}{\p x^{2}}+\frac 12 \g U_{b}\Lbd \{\Phi\}.
\label{syv1}
\ee
Here,
$ \Phi $ is the interface of a distorted planar flame,
$ U_{b} $ is the speed of the planar flame relative to the burning gas,
$  D_{M} $ is the Markstein diffusivity and
$ \g $ is the thermal expansion coefficient,
\be
\gamma=\frac{(\rho_{f}-\rho_{b})}{\rho_{f}},
\label{gamma}
\ee
where
$ \rho_{f} $ is the density of the fresh mixture and
$ \rho_{b} $ is the density of the burned gas,
$ \rho_{f}> \rho_{b}. $ The equation (\ref{syv1}) is
asymptotically exact in the limit of small
$ \gamma \ll 1. $  $\Lbd\{\ldots \}$ represents a linear singular
nonlocal operator defined  conveniently in terms of the spatial
Fourier transform by:
\be
\Lbd: \tilde\Phi (k,t)\longmapsto 2\pi |k|\tilde\Phi (k,t), \quad
\tilde\Phi (k,t)=
\int _{-\infty}^{+\infty}dk {\ }
 \Phi (x,t) e^{2\pi ikx}.
\label{Lbd}
\ee
$ \Lbd $  is responsible for the Darrieus-Landau
instability \cite{2,3}.

Direct numerical simulations for (\ref{syv1}) performed in
\cite{MS} show  that even when the initial conditions is
chosen to be smooth, the cusps develop on the flame interface
as time increases. When the integration domain
is wide enough, the secondary randomlike
subwrinkles  arise on the interface.
 This wrinkling process is  accompanied
by the flame speed enhancement undergoing an  acceleration
in time \cite{ASL}, namely the mean radius seems to increase with
 time according to a $ t^{3/2} $ power-law.

Numerous analytical investigations devoted to  (\ref{syv1})
display that in the limit of long times the local flame
dynamics is driven by the large-scale geometry \cite{13}-\cite{15}.
Exact solutions of  (\ref{syv1}) can be obtained in principle
 by using the pole
decomposition  technique \cite{ASL},\cite{tfh}- \cite{b}.
For such pole solutions, (\ref{syv1}) formally reduces to a finite
set of ordinary  differential equations (ODE) which describe the
motion of the poles in the complex plane. These poles  are
interpreted as to be related to the cusps observed in physical space.
However, numerical and analytical results
demonstrate convincingly that the solutions of the
ODEs do not resemble those obtained from the direct numerical
integration of (\ref{syv1}), \cite{ASL}. In particular, the number
of wrinkles obtained from the ODEs is independent of time (see \cite{tfh})
and the corresponding (mean) expansion of the front is much slower than
the $ t^{3/2} $ power law.

In \cite{pre96}-\cite{ASL}, it was argued that the  inconsistencies with the
pole decomposition method lie in the stability of the exact pole
solutions. The initial value problem of the linearised PDE about a pole
solution  has been solved  numerically, as a result they concluded that
pole solutions are unstable for large
$ \g $.  Consequently, they
are not observed in experiments.

 It was  conjectured in \cite{Joul} that nonlinearity
 alone is not enough to meet the experimental observations and that
the results of the spectral numerical integrations is due to
computational noise. In \cite{Joul} a model had been
developed, where pseudorandom forcing is included. It is shown that
many broad-banded exciting fields indeed lead to the rapid spawning
of wrinkles.

Analyzing the Fourier spectrum of the
solution of PDE obtained numerically in \cite{ASL}, one observes
that it undergoes the local kinks at very short time.
Occurring in the amplitude of Fourier modes and its harmonics,
these  kinks, are followed by
an oscillation of this amplitude which corresponds, in physical space,
to the birth of new pairs of wrinkles.

The linear stability of the pole decomposition solutions was discussed
in \cite{VM} in details. The exact analytical expressions for the
eigenvalues and eigenfunctions  have been constructed. Based on these
expressions, in \cite{VM}, they demonstrate that for any value of the
parameter
$ \g $ there exists the only asymptotically stable solution with
the largest possible (for this particular value of
$ \g $) number of poles, $N_{\g}$. As the parameter
$ \g $ increases, the equilibrium states of the PDE undergo
a cascade of bifurcations. In this way the new solution
with
$ N $ poles gains stability while the former one with
$ N-1 $ poles becomes unstable.
However,
the nonlinear stability and dynamics of cusps
still remain an important  open question  within such an
approach,  \cite{VM}.

In the present  paper we continue the  investigation
of the kink type behavior of the expanded flame front which
has been started in \cite{pre96}-\cite{ASL}.
In contrast to the study in \cite{VM}, we keep the value of $\g$
fixed and small and use  a pseudorandom force
as an origin of the spawning of wrinkles.

 We demonstrate that the main contribution to
the statistics of wrinkling flame front
 is given by a coupled field-force configurations -
the {\it instantons}. These configurations are related directly
to a very short time (practically instant) kinks between
metastable {\it groundstates} incident to different numbers of poles.
We should stress that we  address the stochastic problem
based on the Syvashinsky equation (\ref{syv1}), but not the
exact solutions of this equation.

The paper is organized as follows. In Sec. II, we formulate
the stochastic problem for  the equation governing
 the slope function dynamics of advancing flame fronts. These
fronts usually either form fractal objects
with  contorted and ramified appearance or they wrinkle
 producing self-affine
 fractals characterized with some critical exponent, \cite{Kuper}.
 In Sec. III, we
analyse the problem from the point of view of critical phenomena theory.
We demonstrate that the  $ t^{3/2} $ power law observed experimentally
would be  a direct consequence of the general Kolmogorov's scaling with
 the critical dimensions of time
$ \Delta^{K}_{t}=-2/3 $ and velocity
$ \Delta^{K}_{v}=-1/3 $ (the dimension of
$ x $ is taken by definition as
$ \Delta_{x}=-1 $) which are well known in the fully developed
turbulence theory (see \cite{183} for a review).

We must say that in actual problem the renormalization group
technique (RG) (which has proved itself so well in the
fully developed turbulence theory, \cite{183}, \cite{pre98})
is ultimately ineffective since, obviously, the regime of
critical scaling is not attained. One even can hardly use
the concept of critical dimensions for the actual quantities.

The examples of successful application of the saddle point
calculations  to the Burger's equation \cite{M1} and to the
description of  intermittency phenomenon in turbulence \cite{M2}
have been given recently. These papers have inspired us to employ
this technique in the problem of wrinkling flame front expansion.
The infinite set of instanton like solutions which we have found
is dramatically dissimilar to those computed in \cite{M1} and \cite{M2}.

In Sec. IV, we construct the statistical theory of wrinkles
based on the action functional relevant to the actual stochastic problem.

The minimization of action discussed in Sec. V
 requires that the field and force to be coupled in
some particular configurations.
 We also illustrate the instanton mechanism
of poles generation for the particular two poles initial configuration.
The process keeps repeating itself as time increases. We then conclude
in the last section.

\section{The stochastic problem for the Syvashinsky equation}

 The stochastic
 problem for the equation governing  the slope function dynamics
 of the flame front,
$ u(x,t)=\p_x \Phi(x,t), $ with a Gaussian distributed pseudorandom
force included in the r.h.s. reads as follows
\be
\p_{t} u+ U_{b}u\p_{x} u=
D_{M} \p_{x}^2 u+\frac 12 \gamma U_{b}\Lbd \{u\} +f.
\label{syv2}
\ee
The pair correlation function for
$ f  $ is taken in the form
\be
\langle f(x,t)f(x',t')\rangle =D_{f}(x-x')\delta(t-t'),
\label{c}
\ee
in which the function
$ D_{f}(x-x') $ is supposed to be an even smooth "bell"-shaped
function of
$ x. $ To be specific, we take it in the form
\be
D_{f}(x)=\frac{D_{0}}{\pi}\frac{m}{x^{2}+m^{2}}
\label{dd}
\ee
decaying at the rate
$ m $ and turning into
$ D_{0}\delta(x) $ as
$ m\to 0, $ where
$ D_{0} $ is a constant.

The equation (\ref{syv2}) is similar to the Burger's equation
except the singular term,
$ \propto \gamma U_{b}\Lbd \{u\}.  $
The homogeneous unforced equation (\ref{syv2}) considered
in \cite{tfh} in details. In particular, it was shown that it
possesses a pole decomposition, i.e.,
it allows a countable number
 of uniform  solutions,
\be
u(x,t)=-2 \nu \sum^{N}_{i=-N}\frac 1{x-z_{i}(t)},
\label{pd}
\ee
in which
$ z_{i}' $s are poles in the complex plane (coming in complex
conjugate  pairs)
moving according to the laws of motion of poles
\be
\dot z_{i}= - 2\nu \sum _{i\ne j}\frac 1{z_{i}-z_{j}} -
i\g U_{b}{\ } sign\left[{\Im}(z_{i})\right],
\label{poles}
\ee
where
$ {\Im} $ denotes the imaginary part of a pole.
One can derive easily the corresponding steady
($\dot z_{i}=0$),
solution of the Sivashinsky equation for the simplest
configurations concerning the minimal number of poles.
For example, for two poles the only steady solution  is given
by
\be
u^{(2)}(x)=- \frac{4D_M x}{x^{2}+D_M^2},
\label{1i}
\ee
and there are  two possible four-pole steady configurations,
\be
u^{(4)}(x)=\pm \frac{4D_M (\pm 2x^{3}+27\sqrt{2}iD_M^3+9\sqrt{2}iD_M x^{2})}
{-x^{4}\pm 54\sqrt{2}iD_M^{3} x \pm 6\sqrt{2}iD_M x^{3}+81D_M^4}.
\label{4poles}
\ee
In \cite{pre96}-\cite{ASL}, by numerically solving the
initial value problem that results from linearizing the PDE
about a pole solution, it was concluded that such pole solutions
with a fixed number of poles are asymptotically
unstable, while the actual solutions which can be
observed  experimentally  undergo a sequence of
kinks between different metastable
 {\it groundstates} of the type (\ref{pd}).

To construct the solutions with spawning wrinkles, we exploit
the  exact correspondence between an arbitrary stochastic dynamical
problem  with the Gaussian distributed random force and a quantum
field theory, \cite{MSR}.  A short and elegant proof had been given
in \cite{1983}.  The stochastic problem
(\ref{syv2}) and (\ref{c}) is
completely equivalent to the field theory of two fields  with an
 action functional
\be
S[u,w]=\frac 12\int dt{\ } dx {\ } dy {\ } w(x,t) D_{f}(x-y,t)w(y,t) \\
\label{s}
\ee
$$
-i
\int dt{\ } dx {\ } w\left(  u_{t}+ U_{b}uu_{x}-D_{M}u_{xx}
-\frac 12 \gamma U_{b}\Lbd \{u\} \right).
$$
Here,
$ w(x,t) $ is the auxiliary field which comes into play instead
of the random force
$ f. $
$ w(x,t) $ determines the response functions of the system, for instance,
the linear response function is
$ \left< uw\right>. $

\section{ The power law of the flame front expansion}

The mean squared distance of propagating flame front,
$ R^{2}(t), $  can be expressed naturally
via the linear response function mentioned at the end of the
previous section as following,
\be
R^2(t)=\int dx {\ } x^2 \left< u(x,t)w(x,0)\right>.
\label{R^2}
\ee
The requirement that each term of the action functional be
dimensionless (with respect to
$ x $ and
$ t $ separately) leads to the power counting relation for
the product, $\Delta[uw]=d,$ where
$ d $ is the space dimensionality.
Therefore, the power of the linear
response function is
\be
\Delta[\left<uw\right>]=d-d=0.
\label{uw}
\ee
We note that (\ref{uw}) is still valid whether a critical
regime is attained or not.

Following a tradition, we accept
the natural normalization condition that
$ \Delta [x]=-1, $ then
\be
 \Delta[R^{2}]=-2.
\label{dr}
\ee
From the other hand, the observation data show that
\be
R^{2}\propto t^{3}
\label{rt}
\ee
(i.e.,
$ R\propto t^{3/2} $).
Comparing (\ref{dr}) and (\ref{rt}), one concludes that,
in the theory (\ref{s}), the time variable
$ t $ possesses  the effective  dimension,
\be
\Delta[t]=-2/3,
\label{ct}
\ee
which is equal to the  Kolmogorov's critical dimension of
 time in  fully developed turbulence.

If the result (\ref{ct}) were a true critical dimension,
then one could claim that there is a critical regime in
the wrinkling  flame front propagation problem, i.e., for any
correlation function there is a definite stable large scale long time
asymptotics. Nevertheless, we  stress the dramatic difference between
the stochastic theory  of turbulence (the Navier-Stocks equation
with a random forcing included, \cite{183}) and the actual
problem.  From the point of view of the critical phenomena theory,
the problem (\ref{syv2}) - (\ref{c}) is formulated  erroneously.

Namely, interesting in the long time large scale
asymptotics behavior of correlation functions,
one has to omit the term $ \propto D_{M}k^{2} $ (in the
momentum Fourier  space) from the action (\ref{s}) in benefit to
 $ \propto \g U_{b}|k|. $
 However, in this case, there are infinitely many
Green's functions which have singularities with respect to
a general dilatation of variables, i.e., such a theory
cannot be renormalized.

One can investigate a theory in which the both terms are included
simultaneously. Up to our knowledge, a model  where the concurrence
between two terms (in the momentum Fourier space) $ \propto k^{2} $ and
$ \propto |k|^{2-2\alpha} $ ($ 0<\alpha<1/2 $) was considered firstly
in \cite{NU} in the framework of the renormalization group approach.
It is shown that up to the value $ \alpha_{c}<1/2 $ a regular expansion in
$ \alpha $  and $ \varepsilon $ (the deviation of the space dimensionality
from its logarithmical value) can be constructed and then summed over by
the standard renormalization group procedure. The critical indices
of all quantities are still fixed on their kolmogorovian values.

However, for $ \alpha> 1/2 $ and for  the particular case of
$ \alpha_{r} =1/2 $ which we are interested in, the renormalization
group method  fails. The matter is in new additional singularities
 which spawning in the correlation functions of the  field
$ u $ as $ |k|\rightarrow 0 $. Such singularities cannot be handled by
the renormalization group in principle  since they do not related to a
general scaling with respect to dilatation of variables.

The summation of the leading infrared ($ |k|\rightarrow 0 $) singularities
of correlation functions can be done by an infrared perturbation
theory. If we limit ourselves to the functions $ u_{z}(x,t) $ which
 have poles $ \{z(t)\} $ in the complex plane, then the
 action of the singular   operator
$ \Lbd $  is reduced to a first order derivative operator
\be
\Lbd\{u_{z}\}=i {\  } sign\left(\Im[z(t)]\right)\p_x u_{z}(x,t),
\label{Lu}
\ee
see \cite{tfh}.
Then, in the momentum-frequency representation, the term
with
$ \Lbd $
can be taken into account as a small shift of frequency ($\propto \g$),
\be
\omega \to \omega -i\g U_{b}|k|.
\label{wxx}
\ee
The infrared perturbation theory results from the expansion
of
$ \exp S $ over nonlinearities. The corresponding diagram
technique  coincide with the diagram technique of Wyld, \cite{Wy}.
The lines in the diagrams are associated with the bare propagators,
in the Fourier space,
\be
G_{uw}=G^{*}_{wu}=   \frac 1{-i(\omega-i\g U_{b}|k|)+D_{M}k^{2}}
\label{uw0}
\ee
and
\be
G_{uu}=  \frac 1{-i(\omega-i\g U_{b}|k|)+D_{M}k^{2}} D_{f}(k)
\frac 1{i(\omega-i\g U_{b}|k|)+D_{M}k^{2}},
\label{uu0}
\ee
where
$ D_{f}(k) $ is the momentum representation of (\ref{dd}).
For any correlation function, this diagram technique gives an
infrared representation  which is naturally consistent with
the $ \g $ expansion and is well defined for small values of
the parameter $ \g $.

We are not going to discuss the application of the infrared
perturbation theory to the actual problem in detail.
Here, we conclude that for any pseudo-differential operator
$\propto |k|^{2-2\alpha}, $
$0< \alpha< 1/2, $ the model of the type (\ref{s}) has a
critical regime with the critical indices fixed at their
Kolmogorov's values (see \cite{NU}). However, for
$ \alpha\geq 1/2 $ the stability of asymptotics
 still an important open question if $ \g $ is large.

\section{Statistics of planar flame front wrinkling}

We are going to discuss the saddle point configurations of (\ref{s})
which can provide us by a detailed description of mechanism of
wrinkles generation on the propagating flame front surface.
For the  future purposes, it would be convenient to  perform consequently
  the rescaling of  fields in (\ref{s}),
\be
u \rightarrow \frac{u}{U_{b}},\quad w\rightarrow U_{b} w,
\label{rs1}
\ee
such that the parameter
$ U_{b} $ is removed from the nonlinear term
$ wuu_{x} $, and then another rescaling,
\be
u \rightarrow \frac{u}{\gamma},\quad U_{b}\rightarrow \frac {U_{b}}{\gamma}.
\label{rs2}
\ee
As a result of such a simple transformation, we observe that the parameter
of thermal expansion
$ \gamma $ which we assume to be small, plays the formal role of
$ \hbar $ in quantum field theory:
\be
S\rightarrow \frac 1{\gamma} S.
\label{ss}
\ee
The correlation functions of the basic field
$ u $ are then given by the functional integral
\be
\label{cf-u}
G_{n}(x_{1},t_{1};x_{2},t_{2};\ldots x_{n},t_{n})=\int {\cal D}u {\ } {\cal D}w {\ }
u(x_{1},t_{1})u(x_{2},t_{2})\ldots u(x_{n},t_{n})\exp (-\frac 1{\g} S),
\ee
and can be  derived naturally by means of a generating functional
which has been introduced firstly in \cite{DD} and
then employed in \cite{M1,M2}
\be
{\cal Z}(\lbd) \equiv \left< \exp \left( i\int dt {\ } dx {\ }
\lbd u \right) \right> =\int {\cal D}u{\ }{\cal D}w{\ }
\exp\left(\frac 1\gamma\{-S+i\int dt {\ } dx {\ }\lbd u\}\right).
\label{gf}
\ee
The coefficients of the expansion of
$ {\cal Z} $ in   $\lbd $   are the correlation functions (\ref{cf-u}).

There are no general methods to compute such a  functional integral
exactly. The straightforward perturbative approach is to expand
the exponential in the functional integral (\ref{gf}) in powers of the
nonlinear term
$ wuu_{x}. $ However, since  we are interested in
nonperturbative effects,  it seems more natural to
search  for some saddle-point configurations
which minimize the action functional (\ref{s}),
thus dominating the functional
 integral in a way similar to the saddle point approximation
in ordinary integrals.
Such solutions are called  {\it instantons}, and they determine
the asymptotics of (\ref{gf}) at small
$ \gamma \ll 1 $ which
corresponds to WKB approximation in quantum field theory ($\hbar  \ll 1$).

  Another quantity which can be expressed via the generating functional
(\ref{gf}) is the probability distribution function
$ {\cal P} (u) $ for the field
$ u, $
\be
{\cal P}(u)=\int {\cal D}\lbd {\ } {\cal Z}(\lbd) \exp
 \left(-i\int dt {\ } dx {\ } \lbd u\right).
\label{pdf}
\ee
The behavior of
$ {\cal P} (u) $ for large
$ u $   is also dominated by some saddle-point configurations
of the integrand. However, these configurations are not the same for both
(\ref{gf}) and (\ref{pdf}).

\section{ Kink solutions of the forced Syvashinsky equation}

In what follows we shall look for  saddle point configurations driven
by the random force
in terms of  functions which have poles in the complex plane,
$ u_{z}(x,t). $
To be specific, we observe  generation of
the four poles configuration from the two poles
 as a result of a kink.
In contrast with \cite{M1} and \cite{M2}, we need not introduce
here a large artificial  parameter to fix the
 saddle points dominating  the functional
integrals  (\ref{gf}) and (\ref{pdf}) since we have
the inverse thermal expansion coefficient
$ 1/\g $ which is naturally  large.

Suggesting that the field
$ u $ can be continued analytically on the complex plane
except for the poles,
we shall study the correlation function
of the form
\be
G(z)=\left< \exp \left[\frac {
u(z) -u(z^{*})
 }{\g} \right]\right>
\label{2pcf}
\ee
of two distinct points of the complex plane
 symmetrical  with respect to the real axis.
We suppose also that at the initial moment of time
$ t=0 $ the field
$ u $ can be depicted as a  configuration of two complex conjugated poles,
$ z $ and
$ z^{*}. $
The function (\ref{2pcf}) possesses a generating property:
 Tailoring (\ref{2pcf}) in powers of
$ 1/\g, $ one obtains the "structure functions" for the
 field $ u. $ The functional Fourier transform (\ref{pdf})
of (\ref{2pcf}) gives us the two  point probability distribution.
The structure function generated by (\ref{2pcf}) is related to
the same point
$ x=\Re(z) $ on the real axis.

 Taking an average in (\ref{2pcf}) with respect to
a functional measure, we perform an integration over all possible
configurations
$ u(x,t) $  with the asymptote prescribed by initial two poles
 and all possible final multipole configurations.
The  basic symmetry of the action (\ref{s}) is the Galilean invariance
which reveals itself in the real  transformation
\be
u_{a}(x,t) \longmapsto u(x+X_{a}(t),t) -a(t),
\label{gi}
\ee
where
$ a(t) $ is an arbitrary function of
$ t $ decreasing rapidly as
$ |t|\to \infty ,$ and $ X_{a}(t)=\int_{0}^{t}dt' {\ } a(t'). $
The transformation (\ref{gi}) defines an orbit
in the functional space of
$ u $ along which the result of functional averaging does not change.
It follows that the integral itself is proportional to the
volume of this orbit. This volume should be factorized before one
can perform the saddle-point calculation (see \cite{fp}). It is appropriate
to choose for the latter the "plane" transversal to the real axis
$ \Re(z) $ and then cancel out the real components of
$ u $ which are related to each other via (\ref{gi}).
 In (\ref{2pcf}), the real contribution to
$ u $ is subtracted out, so that  it is very suitable for
 instanton calculations.

The asymptotics of small
$ \g $ in (\ref{2pcf}) is dominated by the saddle point configurations of the
functional
\begin{equation}
{\cal W}[u,w,z]=\frac
{u(z) -u(z^{*})}{\g}
-S[u,w],
\label{w1}
\end{equation}
which  should satisfy the following equations
obtained by varying of (\ref{w1})
with respect to
$ u $ and
$ w $:
\be
u_{t}+ uu_{x}-D_{M}u_{xx}-\frac 12  U_{b}\Lbd \{u\}=
-\frac 12 \frac{iU_{b}}{\gamma} \int  dx'{\ }
 D_{f}(x-x',t)w(y,t),
\label{eq1}
\ee
\be
w_{t}+ uw_{x}+D_{M}w_{xx}+\frac 12  U_{b}\Lbd \{w\}=-\frac{i}{\g}\delta(t)
\left\{
\delta(x-z)-
\delta(x-z^{*})
\right\}.
\label{eq2}
\ee
These equations for the saddle point configurations
 are similar to those derived in \cite{M1}
except the last  singular term in the l.h.s. They follow from the
Syvashinsky equation for the slope function (\ref{syv2}), however
they contain  information on a special force configuration necessary
to produce  instantons also.

Indeed, the particular solutions of (\ref{eq1}) and (\ref{eq2}) are
dependent substantially from the  initial data
 for
$ u $ and
$ w. $
Minimization of the action  requires
$ u\rightarrow 0, $ at
$ t\rightarrow -\infty $ and
$ w\rightarrow 0, $  at $ t\rightarrow \infty. $
Obviously, any solution of the equation (\ref{eq2})
 which is nonsingular  as
$ t\rightarrow +\infty $  should be equal to  zero at
$ t>0 $ (since the field
$ w $ feels  a negative diffusivity). Following an analogy
 with \cite{M1,M2}, one can say that
the field $ w $  propagates backwards in time starting from
its initial value
\be
w(t=-0)=
-\frac{i}{\g_{0}}\left\{
\delta(x-z(0))-
\delta(x-z^{*}(0))
\right\}
\label{iv}
\ee
while it is zero at all later moments of time.
Therefore, the system (\ref{eq1})  and (\ref{eq2}) as well as
the integrals in (\ref{s}) can be treated for
$ t<0 $ only.

While propagating backward in time, (\ref{iv}) is a subject to
a drift of the initial conditions as governed by the velocity
in the equation (\ref{eq2}), the smearing of the initial
$ \delta-$function distributions in (\ref{iv}) due to diffusivity,
and finally, an advection in the complex plane, in the imaginary
direction towards the real axis. In the  limit of no diffusivity,
one can neglect the smearing in the equation (\ref{eq2}).
A simplified equation, which we arrive at when we drop the diffusivity
term is just moving the $ \delta-$singular right hand side  of (\ref{eq2})
around.
Therefore, the solution of (\ref{eq2}) can be expressed naturally
in the form
\be
w(t)=
-\frac{i}{\g(t)}\left\{
\delta(x-z(t))-
\delta(x-z^{*}(t))
\right\}
\label{wd}
\ee
with the boundary conditions
$\g(0)=\g_0,\quad z(0)=z_{0}.$
If one takes the diffusivity term in (\ref{eq2}) into account
(in the case of eventually small
$  D_{M}/\g U_{b}\leq 1 $),
the solution of (\ref{eq2}) would be expressed in terms of even  functions
decaying as
$| x|\to \infty, $
\be
w(t)=
-\frac{i}{\pi\g(t)}\left\{
\frac{y(t)}{y^{2}(t)+(x-z(t))^{2}}-
\frac{y(t)}{y^{2}(t)+(x-z^{*}(t))^{2}}
\right\},
\label{wdx}
\ee
where the parameter
\be
 y(t)=\frac {D_{M}}{\g (t)U_{b}}
\label{y}
\ee
 is a size of an  instanton,  and
$ z(t) $ is its position changing with time
(we have borrowed the terminology from the quantum field theory).
As
$| y(t)|\to 0 $, the instanton  shrinks to a point,
and the solution (\ref{wdx}) is reduced  to (\ref{wd}).

To proceed with the equation (\ref{eq1}), we rewrite it due to
(\ref{Lu}) in the form
\be
u_{t}+ uu_{x}-D_{M}u_{xx}-\frac i2  U_{b}
  sign{\ }\Im[z(t)] \p_x u_{z}(x,t)
= -\frac 12 \frac{iU_{b}}{\gamma} \int  dx'{\ }
 D_{f}(x-x',t)w(y,t).
\label{eq11}
\ee
The one-dimensional  equation (\ref{eq11}) can be linearised
by the Cole-Hopf
transformation,
\be
u(x,t)= - 2D_{M}\frac{\psi_{x}(x,t)}{\psi(x,t)},
\label{ch}
\ee
so that we arrive at the equation
\be
\psi_{t}-D_{M}\psi_{xx}-\frac i2  U_{b}
  sign {\ }\Im[z(t)] \p_x \psi(x,t)
= -\frac 12 \frac{iU_{b}}{\gamma_{0}} \int  dy{\ }
 D_{f}(x-y,t)w(y,t).
\label{neq}
\ee
Since $ w(x,t) $ distinguishes from zero only for
$ t\leq 0 $ and
$ D_{f}\propto \delta(t) $, the only nontrivial
contribution into the r.h.s. of (\ref{neq}) is
given by the moment of time
$ t=0. $
The general solution of the equation (\ref{neq})
in the Fourier space reads as  follows
\be
\psi_{inst}(k,t)= \frac 12 \frac{iU_{b}}{\gamma_{0}}
 D_{f}(k,0)w(k,0)\delta(t),
\label{sol}
\ee
plus a transient process decaying rapidly  as time growing,
$ \propto \exp\left[-t(D_{M} k^{2}+U_{b}|k|)\right].$

Now we can use (\ref{dd}) and (\ref{wdx}) to write down
the r.h.s. of (\ref{sol})  explicitly,
\be
D_{f}(k,0)w(k,0)=4\pi^2 i D_{0}\exp\left[-(y(0)+m)|k|-ik \Re[z_{0}]\right]
\sin k \Im[z_{0}].
\label{rhs}
\ee
Performing an inverse Fourier transform of (\ref{sol}),
 one obtains
\be
\psi_{inst}=\frac{D_{0}U_{b}^{3}\g_0}{2\pi } \left[
\frac 1{(D_{M}+ \g_0U_{b}m)^{2}+\g_0^{2}U_{b}^{2}(x-z_{0})^{2}}-
\frac 1{(D_{M}+ \g_0U_{b}m)^{2}+\g_0^{2}U_{b}^{2}(x-z^{*}_{0})^{2}}
\right].
\label{psiinst}
\ee
Finally, we arrive (\cite{last}) at the following four poles configuration
for the instanton
$ u_{inst}, $
\be
u_{inst}= \frac{4D_{M}x \Im[z]}
{\left((D_{M}/\g_0U_{b}+m)^{2}+x^{2}-\Im[z]^{2}\right)^{2}+
4x^{2}\Im [z]^{2}} \delta(t).
\label{uinst}
\ee
For  the last step of  instanton computation we
have to define the functions
$ \g(t) $ and $\varphi (t)\equiv\Im [z(t)]$  in the equation (\ref{wdx})
(remember that
$ \Re[z] $ is fixed) as
$ t<0. $
We can do it by a direct substitution of (\ref{wd})
(in case of  eventually small diffusivity) into the
equation (\ref{eq2}). Here we note that since the instanton
solution (\ref{uinst}) exists as $ t\geq 0 $
 (if one takes the transient process into account),
 we have  to use the initial two pole configuration
instead of
$ u(x,t) $ in (\ref{eq2}).
As a result,  we obtain the system of simplified equations
\be
-\dot \g(t) =\frac{4xD_{M}}{x^{2}+\varphi(t)^{2}}\g(t),
\label{eq3}
\ee
\be
-\dot \varphi(t)= \frac{4xD_{M}}{x^{2}+\varphi(t)^{2}} \varphi(t).
\label{eq4}
\ee
The formal solution of (\ref{eq3}) is given by
\be
\g(t)=\g_0\exp \left(-\int^{0}_{t}
\frac{4xD_{M}}{x^{2}+\varphi(t')^{2}} dt'\right), \quad (t'<0)
\label{sol3}
\ee
and can be computed, in principle, if one knows
$ \varphi(t).$ The equation (\ref{eq4}) is equivalent to
\be
\frac {x}{D_{M}}\ln \varphi(t)+\frac 1{2x D_{M}}\varphi^{2}(t)+t=C,
\label{eq5}
\ee
which leads to
\be
\varphi(t)=\varphi_{0}\exp \left[-\frac {tD_{M}}{x}-
\frac 12 W\left( \frac {e^{-2 tD_{M}/x}}{x^{2}}\right)\right].
\label{eq6}
\ee
$ W(x) $ is the Lambert function which meets the equation
\be
W(x)\exp W(x)=x.
\label{Lam}
\ee
The latter equation has an infinite number of solutions
 for each (non-zero) value of $x$.
$W$ has an infinite number of
  branches numbered by an integer number
$ n\in [-\infty\ldots\infty] $. Exactly one of
these branches is analytic at 0
(the principal branch,
$ n=0 $).
 The other branches all have a branch
   point at $0$. The principal branch
is real-valued for $x$ in the
    range $ -\exp(-1) \ldots \infty$, while the
image of $-\infty \ldots -\exp(-1)$
    under $W(x)$ is the curve $-y \cot(y) + yi$, for $ y \in [0 \ldots \pi]$.
For all the branches other than the principal branch, the branch cut
 dividing them is the negative real axis.
The image of the negative real axis under the branch
$W(n,x)$ is the curve $ -y \cot(y) + yi$,
 for $ y \in [2k\pi \dots (2k+1)\pi$ if $k > 0$ and $ y \in
 [(2k+1)\pi \ldots (2k+2)\pi]$
if $ k < -1$. These curves, therefore, bound the ranges of the branches
of $W$, and in each case, the upper boundary of the region is
 included in the range of the corresponding branch.

Each particular orbit of (\ref{eq6}) provides a distinct solution
of (\ref{sol3}) and (\ref{wd}). However, each configuration
$ w(x,t), $
$ t<0 $ which enjoys (\ref{wd}) is related to the same
configuration
$ u(x,t) $,
$ t>0. $ The value of  (\ref{w1}) for the
instanton is obviously finite, however, one can hardly compute it
for each branch of
$ w(z,t). $

Let us consider   the principle branch of the function
$ \varphi(t) $ just to illustrate the idea of computation.
 One can check  that the leading contribution
to $\varphi$  is accumulated around
$ x=0. $
The asymptotic behavior of $W$ at complex infinity and at
$ 0$  is given by
\be
W(x) \sim \log(x)  -\log(\log(x))+
  \sum_{m,n=0}^{\infty}C(m,n)\frac{\log\left(\log(x)\right)^{(m+1)}}
{\log(x)^{(m+n+1)}},
\label{Las}
\ee
where
$ \log(x)$ denotes the principal branch of the logarithm,
 and the coefficients  $C(m,n)$ are constants.

Restricting to the first term of the asymptote (\ref{Las}),
one obtains
\be
\varphi(t)\simeq - \varphi(0) D_{M}t, \quad (t<0),
\label{vf}
\ee
then we use (\ref{vf})  and (\ref{sol3}) to compute
$ \g(t) $ as
$ t<0, $
\be
 \g\simeq \g_0
\exp[\frac{4}{\varphi_{0}}\tan^{-1}\frac{\varphi_{0}D_{M}t}{x}], \quad t<0.
\label{gam}
\ee
Now it is a matter of a simple computation to find
the action on the principal branch of the instanton.
We collect everything together
and substitute (\ref{gam}), (\ref{vf}), (\ref{uinst}),
and (\ref{wd}) back to (\ref{s}) to obtain
\be
S_{inst}=-\frac {D_{0}}{m}\exp[-\frac{2}{\varphi_{0}}\tan^{-1}\varphi_{0}],
\label{sinst}
\ee
while the correlation function we have been studying is
\be
G\propto \exp \left(
\frac {D_{0}}{m}\exp[-\frac{2}{\varphi_{0}}\tan^{-1}\varphi_{0}] \right).
\label{cfinst}
\ee
This simplified formula is obviously not an exact answer. It is just  a
leading asymptotic of
$ G(\Im[z_{0}]) $ if
$ 1/\g_0 $ is a large number,
$ \varphi_{0}\equiv \Im[z_{0}] $ is eventually small, however,
not too small (since we have not  taken the smearing due to
diffusivity into account). Nevertheless, for some interval
of scales $ \Im[z_{0}] $, rather small than large (see Fig. 1),
  the asymptotics prescribed by (\ref{cfinst})
(a thick line) is very close to
a model curve which corresponds to the Kolmogorov's
critical exponent for  velocity, $ \Delta^{K}[v]=-1/3$.                                    .

Due to a specific property of the action (\ref{s}),
the contribution from fluctuations of
$ \delta w $  and
$ \delta u $ up to  second order  will be zero.
To find higher order corrections to (\ref{cfinst}),
one has to consider at least the third order terms.
We investigate this problem in future publications.

\section{Discussion and Conclusions}

The study performed in this paper confirms that the stochastic
model for the outward propagating  flame  in the regime of well
developed hydrodynamic instability leads to the self-fractalization of the
flame front. The mechanism consists of successive instabilities
through which the interface becomes more and more wrinkled as time
increases. The main contribution to the statistics of wrinkles
is due to a coupled field-force configurations which are found to be
responsible for the birth and growth of wrinkles. This behavior
is different from exact pole solutions for which the
number of poles is constant. The acceleration of the mean front radius
is clearly due to successive births of poles.

We have demonstrated that the  $ t^{3/2} $ power law observed
experimentally for the mean radius of the advancing flame front
would be  a direct consequence of the  Kolmogorov's scaling with
 the critical dimensions of time
$ \Delta^{K}_{t}=-2/3 $ which is well known in the fully developed
turbulence theory.
Provided the critical regime in the model of wrinkling flame front
 exists, then it  means that each correlation function in the
model has a definite stable long time large scale asymptotics, which does
not depend from the particular sequence of kinks.
If one replaces the nonlocal operator
$ \Lbd $ with some  pseudo-differential operator
$\propto |k|^{2-2\alpha}, $
$0< \alpha< 1/2, $ then, as it was shown in \cite{NU},
 the model of the type (\ref{s}) has a
critical regime with the critical indices fixed at their
Kolmogorov's values.  However, for  $ \alpha\geq 1/2 $ the
stability of asymptotics still an important open question
if $ \g $ is large.

We have used the saddle point calculations
assuming the inverse thermal expansion coefficient
$ 1/\g $ as a large parameter. As a result, we construct an infinite
  family of instanton solutions numbered by
$n\in [-\infty\ldots\infty]$. Each instanton
determined by one of the branches of the Lambert function
$ W(x) $ has a unique behavior as
$ t<0 $ ,
 however, all  instantons are undistinguishable as
$ t>0. $
The asymptotic behavior of these solutions is close
to that of prescribed by the Kolmogorov's scaling.

The crucial problem of the developed technique is
of contribution to the action from the fluctuations against
the instanton background. The general analysis of the
set of instantons which we have found will be
published in a forthcoming paper.

\section{Acknowledgments}

We would like to thank E. Khanin,  M. Nalimov, and M. Komarova for their
valuable advices and discussions.


\begin{thebibliography}{99}
\bibitem{last}
To compute the inverse Fourier transform, we have used
the fact that for any
$ \xi>0, $
$$
  e^{-\xi}=\int^{\infty}_{0}\frac{e^{-\mu}}{\sqrt{\pi \mu}}e^{-\xi^2/4\mu}
d\mu.
$$
\end{thebibliography}
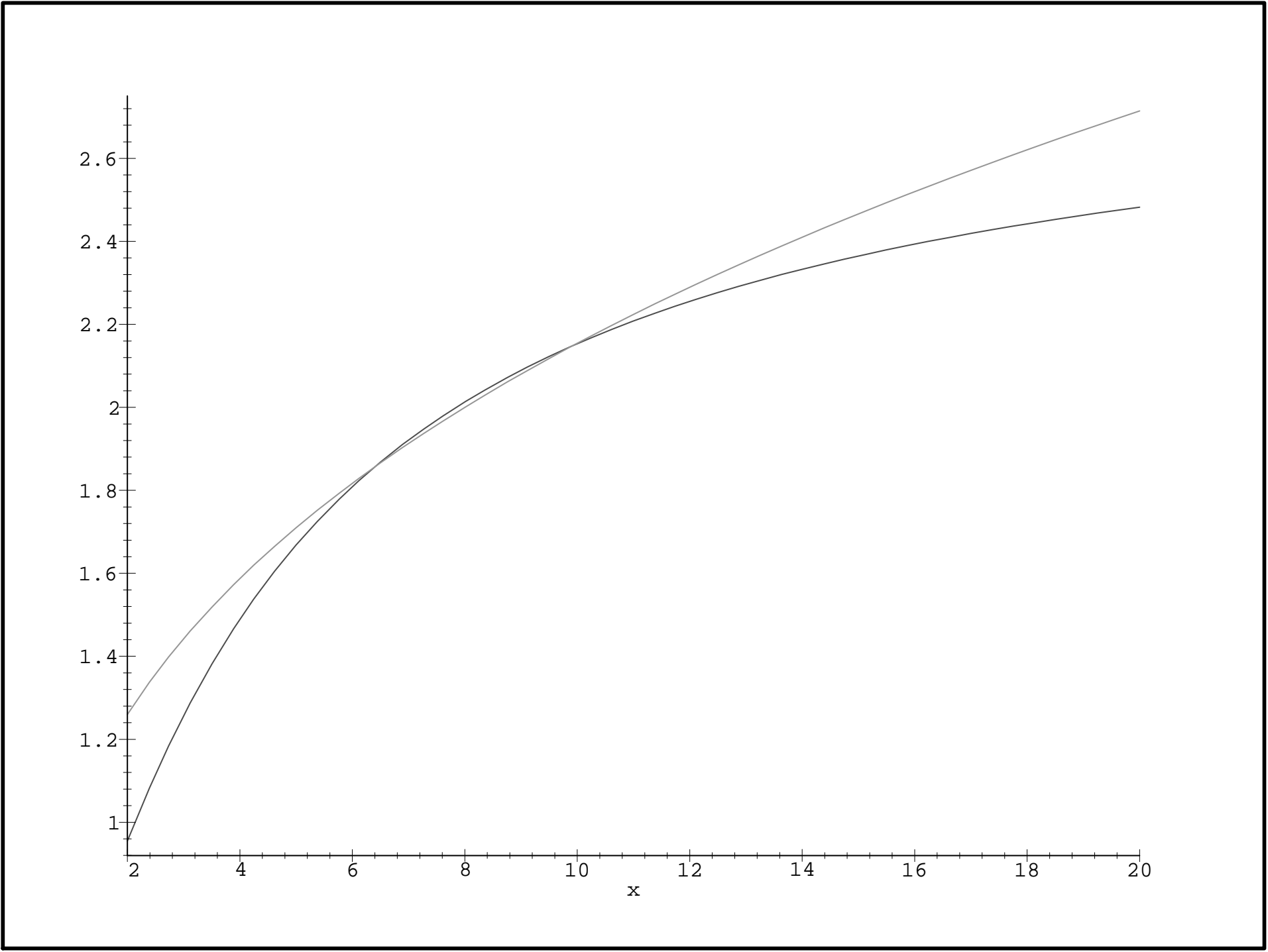
\end{document}